\documentclass[preprint,aps,nofootinbib]{revtex4}

\def\gsim{\begin{array}{c} > \\ \sim \end{array}}
\def\lsim{\begin{array}{c} < \\ \sim \end{array}}

\begin{document}
\draft
\begin{titlepage}
\title{\large \bf Spacetime Foam}\thanks{This essay received an honorable
mention in the Annual Essay
Competition of the Gravity Research Foundation for the year 2002}
\author{Y. Jack Ng}
 \email{yjng@physics.unc.edu}
\address{Institute of Field Physics, Department of Physics and Astronomy,\\
University of North Carolina, Chapel Hill, NC 27599-3255, USA\\}
\bigskip

\begin{abstract}

Spacetime is composed of a fluctuating arrangement of bubbles or loops
called spacetime foam, or quantum foam.  We use the holographic
principle to deduce its structure, and show that the result is
consistent with gedanken experiments involving spacetime
measurements.
We propose to use laser-based atom interferometry techniques
to look for spacetime fluctuations.
Our analysis makes it clear that the physics of quantum foam
is inextricably
linked to that of black holes.
A negative experimental result, therefore, might have non-trivial
ramifications for semiclassical gravity and black hole physics.

\end{abstract}

\maketitle
\end{titlepage}

\newpage

Spacetime appears smooth on large scales.  On small scales, however, it is
bubbly and foamy due to quantum fluctuations.  In this essay, we use the
holographic principle to show that the fluctuations are much larger
than what conventional wisdom leads us to believe.  We
alternatively derive the same results by carrying out
gedanken experiments to measure distances and time intervals.
Intriguingly, the fluctuations are large enough that they may
one day be detectable with improved modern laser-based atom interferometers.

\bigskip

\begin{center}

{\bf From the holographic principle to spacetime foam}\\
\end{center}

The holographic principle grew out of
the profound insights of Wheeler, Bekenstein, Hawking, 't Hooft, and
Susskind.\cite{wbhts}  It states that the maximum number of degrees of
freedom that can be put into a region of space is given by the area of the
region in Planck units.  To connect it to quantum foam, let us consider a
region of space measuring $R \times R \times R$, and imgine partitioning it
into cubes as small as physical laws allow.  With each small cube we
associate one degree of freedom.   If the smallest uncertainty in
measuring a distance $R$ is $\delta R$, in other words, if the fluctuation
in distance $R$ is $\delta R$, then the smallest such cubes have volume
$(\delta R)^3$.  (Otherwise, one could divide $R$ into units each measuring
less than $\delta R$, and by counting the number of such units in $R$, one
would be able to measure $R$ to within an uncertainty smaller than
$\delta R$.)  Thus the maximum number of degrees of freedom,
given by the number of small cubes we can put into the region of
space, is $(R/ \delta R)^3$.  The holographic principle demands that
$(R / \delta R)^3 \lsim (R / l_P)^2$,
where $l_P = c t_P \equiv (\hbar G/c^3)^{1/2}$ is the Planck length.  This
yields
\begin{equation}
\delta R \gsim (R l_P^2)^{1/3} = l_P \left(\frac{R}{l_P}\right)^{1/3}.
\label{nvd1}
\end{equation}
Thus quantum fluctuations from individual bubbles of spacetime
inside a distance $R$ add together
to produce a (curious) $\sqrt[3]{R}$-dependence and are much larger than
the folklore\cite{MTW}
indicates (viz., $\delta R \gsim l_P$).\footnote{Naively, one
would assume $\delta R$ to be given by
$l_P$.  Then the number of degrees of freedom would be given by the
volume of the region of space $R^3$ in Planck units.  This (wrong) choice
of $\delta R$ violates the holographic principle.}
The corresponding metric fluctuation is given by
$\delta g_{\mu \nu} \gsim (l_P / R )^{2/3}$.
Note that even for a
macroscopic distance $R$, the fluctuation
$\delta R$, though much larger than the Planck scale $l_P$, is still
incredibly small; e.g., for $R$ = 1 km, $\delta R$ is to an atom as an atom
is to a human being.  Since the holographic principle is deeply rooted in
black hole physics, this way of deriving spacetime fluctuations is
highly suggestive of the deep connection between quantum foam and black
hole physics.

\smallskip

\begin{center}
{\bf From spacetime measurements to spacetime foam}\\
\end{center}

For an alternative means of deriving $\delta R$, let us consider
a gedanken experiment to measure the
distance $R$ between two points.
The need for carrying out explicit measurements to determine
distances is implicit in
general relativity, according to which, coordinates do not have any
meaning independent of observations; in fact, a coordinate system is
defined only by explicitly carrying out spacetime distance measurements.
Following Wigner\cite{wigner}, we can put
a clock at one of the points and a mirror at the other.  By sending a
light signal from the clock to the mirror in a timing experiment, we can
determine the distance.
However, the quantum uncertainty in the positions of
the clock and the mirror introduces an inaccuracy $\delta R$ in the
distance measurement.  Let us concentrate on the clock and denote its mass
by $m$.  Wigner argued that if it has a linear
spread $\delta R$ when the light signal leaves the clock, then its position
spread grows to $\delta R(2R/c) = \delta R + \hbar R (mc \delta R)^{-1}$
when the light signal returns to the clock, with the minimum at
$\delta R = (\hbar R/mc)^{1/2}$.  Hence one concludes that
\begin{equation}
\delta R^2 \gsim \frac{\hbar R}{mc}.
\label{sw}
\end{equation}
One can supplement this quantum mechanical relation with a limit from
general relativity\cite{ngvan1}.  To see this, let the clock be a
light-clock consisting of two parallel mirrors (each of mass $m/2$), a
distance $l$ apart, between which bounces a beam of light.  For
the uncertainty in distance measurement not to be greater than $\delta R$,
the clock must
tick off time fast enough that $l/c \lsim \delta R /c$.  But $l$, the
size of the clock, must be larger than the Schwarzschild radius $Gm/c^2$ of
the mirrors, for otherwise one cannot read the time registered on the
clock.  From these two requirements, it follows that
\begin{equation}
\delta R \gsim \frac{Gm}{c^2},
\label{ngvan}
\end{equation}
the product\footnote{Here we can appreciate the importance of taking
into account the effects of instruments in this gedanken experiment.
Usually when one wants to examine a certain field
(say, an electromagnetic field) one uses instruments that are neutral
(electromagnetically neutral) and massive for, in that case, the effects
of the instruments are negligible.  But here in our gedanken experiment,
the relevant field is the gravitational field.  One cannot have a
gravitationally neutral yet massive set of instruments because the
gravitational charge is equal to the mass according to the principle
of equivalence in general relativity.  Luckily we can now exploit this
equality of the gravitational charge and the inertial mass of the clock
to eliminate the dependence on $m$.}
of which with Eq.~(\ref{sw}) yields precisely the
expression for spacetime fluctuation $\delta R$
given by Eq.~(\ref{nvd1}), obtained by using the holographic
principle.  (Thus one can actually argue
that the holographic principle has its origin in the quantum
fluctuations of spacetime.)  A gedanken experiment to measure a time interval
$T$ gives an analogous expression:
\begin{equation}
\delta T \gsim (T t_P^2)^{1/3}.
\label{deltat}
\end{equation}

There are also uncertainties in energy-momentum measurements.  Such
uncertainties may account for some unexpected features in the high
energy cosmic ray and gamma ray spectra.\cite{jackng}

\smallskip
\begin{center}
{\bf Interrelationship between spacetime foam and black hole physics}\\
\end{center}

It is interesting that an argument, very similar to that
used above to deduce the structure of spacetime foam,
can be applied to discuss the precision and
the lifetime of a clock.\cite{ng}  For a clock of mass $m$, if the
smallest time
interval that it is capable of resolving is $t$ and its total
running time is $T$,
one finds $t^2 \gsim \frac{\hbar T}{mc^2}$,\cite{ng}
the analogue of Eq.~(\ref{sw}), and $t \gsim \frac{Gm}{c^3}$, the analogue
of Eq.~(\ref{ngvan}).\footnote{
One can combine these two expressions to give
$T/ t^3 \lsim t_P^{-2} = \frac{c^5}{\hbar G}$, which relates clock precision
to its lifetime.\cite{ng}  (Note that this new expression is just
Eq.~(\ref{deltat}) with $t$ playing the role of $\delta T$.)}
Now let us apply these two (in-)equalities to a black hole of
mass $m$ used as a clock.  It is reasonable to use
the light travel time across the black hole's horizon as the resolution
time of the clock,
i.e., $t \sim \frac{Gm}{c^3} \equiv t_{BH}$, then one immediately finds that
$T \sim \frac{G^2 m^3}{\hbar c^4} \equiv T_{BH}$,
which is just Hawking's black hole
lifetime!  Thus, if we had not known of black hole evaporation, this
remarkable result would have implied that there is a maximum lifetime (of
this magnitude) for a black hole.
This is another demonstration of the intimate (if, in this case,
indirect) relationship
between quantum foam and black hole physics.

One can also translate the above
clock relations into useful expressions for a simple computer.  The fastest
possible processing frequency is obviously given by $t^{-1}$. Thus
we identify $\nu = t^{-1}$ as the clock rate of the computer, i.e., the
number of operations per bit per unit time.  The identification of the
number $I$ of bits of information in the memory space of a simple computer
is subtler.  Since $T/t$ is the maximum number of steps of
information processing, we make the identification $I = T/t$.\footnote{Using
the clock relation in the preceding footnote
and the identifications of $\nu$ and $I$
in terms of $t$ and $T$, one gets
$I \nu^2 \lsim \frac{c^5}{\hbar G}$.  This expression links together our
concepts of information, gravity, and quantum
uncertainty.\cite{ng}}  Now imagine that we can form
a black hole (of mass $m$) whose initial conditions encode certain
information to be processed.\footnote{It is possible, in
principle, to program black holes
to do computations in such a way that the results of the computation
can be read out of the fluctuations in the apparently thermal
Hawking radiation, if black holes indeed evolve in a unitary fashion as we
believe.\cite{Lloyd}  For a black hole computer, the inequality
$I \nu^2 \lsim \frac{c^5}{\hbar G}$ is saturated; thus one can even
claim that black holes, once they are programmed to do computations,
are the ultimate computers.}  Then the memory space of the black hole
computer has $I = T_{BH}/t_{BH} \sim (m/m_P)^2$, where
$m_P \equiv (\hbar c/G)^{1/2}$ is
the Planck mass.  This gives the number of bits $I$ as the event horizon
area in Planck units, as expected from the identification\cite{wbhts} of a
black hole entropy!  Furthermore, the number of operations per unit time
for a black hole computer is given by $I \nu \sim mc^2/\hbar$, in agreement
with Lloyd's results\cite{Lloyd} for the ultimate physical limits to
computation.  All these results indicate the conceptual interconnections
of the physics underlying simple clocks, simple computers,
black holes, and spacetime foam.

\smallskip
\begin{center}
{\bf Interferometers as detectors of spacetime foam}\\
\end{center}

Now we come to an important question: how do we detect quantum foam, i.e.,
how do we check Eq.~(\ref{nvd1}) and Eq.~(\ref{deltat})
experimentally?  It has been suggested
that modern gravitational-wave interferometers can potentially provide a
way, because the intrinsic foaminess of spacetime gives another source of
noise in the gravitational-wave interferometers that can be highly
constrained.\cite{ACNVD}
Here we propose to use a smaller and simpler experimental setup; and
following a similar analysis by I. Percival\cite{percival}, we
optimistically suggest that laser-based atom interferometry
experiments
may be precise enough in the not-too-distant future to detect spacetime
fluctuations on the scales of quantum gravity at the level given by
Eq.~(\ref{nvd1}) and Eq.~(\ref{deltat}).  In a laser-based atom
interferometer, an atomic beam is split by laser beams
into two coherent wave packets which are kept apart before being
recombined by laser beams.  The phase change of each wave packet
is proportional to the proper time along its path, and so the resulting
interference pattern depends on the time difference between the two paths.
In the absence of spacetime fluctuations, the phase change $\eta$ over a
time interval $T$ is given by $\eta(T) = \Omega T$, where $\Omega \equiv
mc^2/\hbar$ is the quantum angular frequency associated with the mass $m$
of the atom.  Due to spacetime fluctuations (Eq.~(\ref{deltat})), there is
an additonal fluctuating phase $\delta \eta$ given by
\begin{equation}
\delta \eta \sim \frac{(T t_P^2)^{1/3}}{T} \eta = (T t_P^2)^{1/3} \Omega.
\label{interf}
\end{equation}
For example, in 1992, Chu and Kasevich at Stanford University built an
atom interferometer which used sodium atoms ($m \sim 4.5 \times 10^{-26}$
kg), and the two wave packets were kept apart for 0.2 sec.\cite{chu}  For
that experiment, one finds that $\eta (T) \sim 7 \times 10^{24}$ radians and
$\delta \eta \sim 3 \times 10^{-4}$ radians.  Thus one needs a precision of
about 1 part in $10^{29}$ to look for spacetime foam
(through suppression of the interference pattern), compared
with the precision of 1 part in $10^{26}$ that was then achieved.  In other
words, one needs a (mere) thousandfold improvement in noise sensitivity to
detect spacetime fluctuations.  Though the above argument, a variant of
the one given in Ref. \cite{percival}, is necessarily short and perhaps too
simplistic and overtly optimistic, hopefully the conclusion is not too
far off the mark.

\vspace{1.1cm}

In summary, we have combined the general principles of quantum mechanics
and general relativity to address the problem of quantum fluctuations of
spacetime.  A simple application of the holographic principle has shown
that spacetime undergoes much larger quantum fluctuations than one may
expect.  This result is confirmed by gedanken experiments for
spacetime measurements.  We believe that the
Planck scale, so far only a hypothetical extreme regime,
will eventually become a realm that can be approached and measured,
for instance by interferometry techniques.  In this
essay, we have also
highlighted the interconnection between spacetime foam and black hole
physics.  Hence, if future experiments show
that spacetime fluctuates at a level smaller than our prediction
$\delta R \sim \sqrt[3]{R l_P^2}$,
we will know that our current understanding of semiclassical
gravity and black hole physics may need a considerable revision.  We hope
that these arguments are sufficiently convincing to encourage a
determined experimental quest to detect quantum foam, the very
fabric of spacetime, for, as Faraday wrote:

\begin{quote}
\emph{Nothing is too wonderful to be true, if it be consistent with the
laws of nature, and in such things as these, experiment is the best test of
such consistency.}
\end{quote}

\vspace{1.2cm}

This work was supported in part by the US Department of Energy and the
Bahnson Fund of the University of North Carolina.

\end{document}